\documentclass[11pt]{article}
\usepackage{graphicx}
\usepackage{amsfonts}
\usepackage{theorem}
\newtheorem{Lem}{Lemma}[section]
\newtheorem{Def}[Lem]{Definition}
\newtheorem{The}[Lem]{Theorem}
\newtheorem{Prop}[Lem]{Proposition}
\newtheorem{Cor}[Lem]{Corollary}

\newcommand{\qed}{\hbox{\rule{6pt}{6pt}}}

\newcommand{\Tr}{\mathbf{Tr}}

\begin{document}
\title{Trace inequalities in nonextensive statistical mechanics }
\author{Shigeru Furuichi$^1$\footnote{E-mail:furuichi@ed.yama.tus.ac.jp}\\
$^1${\small Department of Electronics and Computer Science,}\\{\small Tokyo University of Science, Yamaguchi, 756-0884, Japan}}
\date{}
\maketitle
{\bf Abstract.} In this short paper, we establish a variational expression of the Tsallis relative entropy. 
In addition, we derive a generalized thermodynamic inequality and a generalized Peierls-Bogoliubov inequality. 
Finally we give a generalized Golden-Thompson inequality.
\vspace{3mm}

{\bf Keywords : } Trace inequalities, thermodynamic inequality, Peierls-Bogoliubov inequality, Golden-Thompson inequality and Tsallis relative entropy 

\vspace{3mm}

{\bf 2000 Mathematics Subject Classification : } 47A63, 94A17, 15A39

\vspace{3mm}

%%%%%%%%%%%%%%%%%%%%%%%%%%%%%%%%%%%%%%%%%%%%%%%%%%%%%%%%%%%%%%%%%%%%%%%%%%%%%%%%%%%%%%%%%%%%%%%%%%%%%%%%%%%%%%%%%%%%%%%%%%%%%%%%%%%%%%%%%%%%%%%%%%%%%%%%%%%%%%%
%%%%%%%%%%%%%%%%%%%%%%%%%%%%%%%%%%%%%%%%%%%%%%%%%%%%%%%%%%%%%%%%%%%%%%%%%%%%%%%%%%%%%%%%%%%%%%%%%%%%%%%%%%%%%%%%%%%%%%%%%%%%%%%%%%%%%%%%%%%%%%%%%%%%%%%%%%%%%%%
%%%%%%%%%%%%%%%%%%%%%%%%%%%%%%%%%%%%%%%%%%%%%%%%%%%%%%%%%%%%%%%%%%%%%%%%%%%%%%%%%%%%%%%%%%%%%%%%%%%%%%%%%%%%%%%%%%%%%%%%%%%%%%%%%%%%%%%%%%%%%%%%%%%%%%%%%%%%%%%
%%%%%%%%%%%%%%%%%%%%%%%%%%%%%%%%%%%%%%%%%%%  Section1  %%%%%%%%%%%%%%%%%%%%%%%%%%%%%%%%%%%%%%%%%%%%%%%%%%%%%%%%%%%%%%%%%%%%%%%%%%%%%%%%%%%%%%%%%%%%%%%%%%%%%%%%
%%%%%%%%%%%%%%%%%%%%%%%%%%%%%%%%%%%%%%%%%%%%%%%%%%%%%%%%%%%%%%%%%%%%%%%%%%%%%%%%%%%%%%%%%%%%%%%%%%%%%%%%%%%%%%%%%%%%%%%%%%%%%%%%%%%%%%%%%%%%%%%%%%%%%%%%%%%%%%%
%%%%%%%%%%%%%%%%%%%%%%%%%%%%%%%%%%%%%%%%%%%%%%%%%%%%%%%%%%%%%%%%%%%%%%%%%%%%%%%%%%%%%%%%%%%%%%%%%%%%%%%%%%%%%%%%%%%%%%%%%%%%%%%%%%%%%%%%%%%%%%%%%%%%%%%%%%%%%%%
%%%%%%%%%%%%%%%%%%%%%%%%%%%%%%%%%%%%%%%%%%%%%%%%%%%%%%%%%%%%%%%%%%%%%%%%%%%%%%%%%%%%%%%%%%%%%%%%%%%%%%%%%%%%%%%%%%%%%%%%%%%%%%%%%%%%%%%%%%%%%%%%%%%%%%%%%%%%%%%
\section{Introduction}

Recently, the matrix trace inequalities in statistical mechanics are studied by Bebiano et.al. in \cite{BPL}.
 Their results are generalized by them in \cite{BPL2} via $\alpha$-power mean.
In addition, the further generalized logarithmic trace inequalities are obtained 
and their convergences are shown via generalized Lie-Trotter formulae in \cite{Furuta}.
Inspired by their works, we generalize the trace inequalities in \cite{BPL} 
by means of a parametric extended logarithmic function $\ln_{\lambda}$ which will be defined below. 
Our generalizations are different from those in \cite{BPL2,Furuta}.
In the sense of our generalization, we give a generalized Golden-Thompson inequality. In addition, we give a related trace inequality as concluding remarks.

We denote $e_{\lambda}^x \equiv (1 + \lambda x)^{\frac{1}{\lambda}}$ and its inverse function $\ln_{\lambda}x \equiv \frac{x^{\lambda} -1}{\lambda}$,
for $\lambda \in (0,1]$ and $x \geq 0$. 
The functions $e_{\lambda}^x $ and $\ln_{\lambda} x$ converge to $e^x$ and $\log x$ as $\lambda \to 0$, respectively.
Note that we have the following relations:
\begin{equation}
 e_{\lambda}^{x+y+\lambda xy} = e_{\lambda}^x e_{\lambda}^y, \,\,\,\,\,
 \ln_{\lambda} xy = \ln_{\lambda}x + \ln_{\lambda} y + \lambda \ln_{\lambda} x \ln_{\lambda} y. \label{nonex01}
\end{equation}
The Tsallis entropy was originally defined in \cite{Tsa} by $- \sum_{i=1}^n p_i^q \ln_{1-q} p_i = \frac{\sum_{i=1}^n (p_i^q - p_i)}{1-q}$ for any nonnegative real number $q$ and
a probability distribution $p_i \equiv p(X=x_i)$ of a given random variable $X$. 
Taking the limit as $q \to 1$, the Tsallis entropy converges to the Shannon entropy $-\sum_{i=1}^n p_i \log p_i$. 
We may regard that the expectation value $E_q(X) = \sum_{i=1}^n p_i^q x_i$ depending on the parameter $q$ is adopted in order to define the Tsallis entropy as $x_i =-\ln_{1-q}p_i$, 
while the usual expectation value $E(X) = \sum_{i=1}^n p_i x_i$ is adopted in order to define the Shannon entropy as $x_i =-\log p_i$.
In the sequel we use the parameter $\lambda \in (0,1]$ insead of $q$. There is a relation between these two parameters such that $q = 1-\lambda$.  

The Tsallis entropy and the Tsallis relative entropy in quantum system (noncommutative system) are defined in the following manner. 
See \cite{FYK,FYK2} for example.
\begin{Def}
The Tsallis entropy is defined by 
$$
S_{\lambda}(\rho) \equiv \frac{\hbox{Tr}[ \rho^{1-\lambda}  - \rho]}{\lambda} = -\hbox{Tr}[\rho^{1-\lambda} \ln_{\lambda} \rho]
$$
for a density operator $\rho$ and $\lambda \in (0,1]$. The Tsallis relative entropy is defined by
$$
D_{\lambda} (\rho \vert \sigma) \equiv \frac{\hbox{Tr}[\rho - \rho^{1-\lambda}\sigma^{\lambda}]}{\lambda} = \hbox{Tr}[\rho^{1-\lambda} (\ln_{\lambda} \rho - \ln_{\lambda} \sigma)]
$$
for density operators $\rho$, $\sigma$  and $\lambda \in (0,1]$.
\end{Def}
The Tsallis entropy and the Tsallis relative entropy converge to the von Neumann entropy $S(\rho) \equiv -\hbox{Tr}[\rho \log \rho]$ 
and the relative entropy $D(\rho \vert \sigma) \equiv \hbox{Tr}[\rho(\log \rho -\log \sigma)]$ as $\lambda \to 0$, respectively. 
See \cite{OP} for details on the theory of quantum entropy.  
Two Tsallis entropies have nonadditivities such that
\begin{equation}  \label{pseudo_Tsallis}
S_{\lambda}(\rho_1\otimes \rho_2) =S_{\lambda}(\rho_1)+S_{\lambda}(\rho_2)+\lambda S_{ \lambda}(\rho_1)S_{\lambda}(\rho_2),
\end{equation}
and
\begin{equation}  \label{pseudo_relative_Tsallis}
D_{ \lambda}  (\rho_1 \otimes \rho_2 \vert \sigma_1 \otimes \sigma_2) =D_{ \lambda}  (\rho_1\vert \sigma_1)+D_{ \lambda}  (\rho_2\vert \sigma_2 )
-\lambda D_{ \lambda}  (\rho_1\vert \sigma_1)D_{ \lambda}  (\rho_2\vert \sigma_2),
\end{equation}
due to the nonadditivity Eq.(\ref{nonex01}) of the function $\ln_{\lambda}$.
Thus the field of the study using these entropies is often called the nonextensive statistical physics 
and many research papers have been published in mainly statistical physics \cite{AO}. 
In addition, for the relative R\'enyi entropy of order $\lambda$ 
$$
R_{\lambda}(\rho\vert \sigma) \equiv \frac{1}{\lambda}\log Tr[\rho^{1-\lambda}\sigma^{\lambda}],\quad \lambda \in (0,1]
$$
we have the following relation between $R_{\lambda}(\rho\vert \sigma)$ and $D_{\lambda}(\rho\vert\sigma)$ :
$$
\lambda D_{\lambda}(\rho\vert\sigma)+\exp[\lambda R_{\lambda}(\rho\vert \sigma)] = 1.
$$
See our previous papers \cite{FYK,FYK2} on the mathematical properties of the Tsallis entropy and the Tsallis relative entropy.

\section{A variational expression of Tsallis relative entropy}

In this section, we derive a variational expression of the Tsallis relative entropy as a parametric extension of that of the relative entropy in Lemma 1.2 of \cite{HP2}.
A variational expression of the relative entropy has been studied in the general setting of von Neumann algebras \cite{Petz2,Kos}. 
In the sequel, we consider $n \times n$ complex matrices in the finite quantum system.
A Hermitian matrix $A$ is called a nonnegative matrix (and denoted by $A \geq 0$) if $\langle x, Ax\rangle \geq 0$ for all $x \in \hbox{C}^n$. 
A nonnegative matrix $A$ is called a positive matrix (and denoted by $A >0$)
if it is invertible. Throughout this paper, $\Tr$ means the usual matrix trace. A nonnegative matrix $A$ is called a density matrix if $\Tr[A]=1$.
In the below, we sometimes relax the condition of the unital trace for the matrices in the definition of the Tsallis relative entropy $D_{\lambda}(\cdot \vert \cdot)$, 
since it is not essential in the mathematical studies of the entropic functionals.
\begin{The}   \label{gen_rela_vari_exp}
For $\lambda \in (0,1]$, we have the following relations.
\begin{itemize}
\item[(1)] If $A$ and $Y$ are nonnegative matrices, then
$$ \ln_{\lambda} \Tr[ e_{\lambda}^{A+\ln_{\lambda}Y }] = \max \left\{ \Tr[X^{1-\lambda}A] -D_{\lambda} (X\vert Y) : X \geq 0, \Tr[X] =1 \right\}. $$
\item[(2)] If $X$ is a positive matrix with $\Tr[X] = 1$ and $B$ is a Hermitian matrix, then
$$  D_{\lambda}(X\vert e_{\lambda}^B)  =\max \left\{ \Tr[X^{1-\lambda} A] - \ln_{\lambda} \Tr[e_{\lambda}^{A+B}] : A \geq 0 \right\}.$$
\end{itemize}
\end{The}

{\it Proof}:
The proof is almost similar to that of Lemma 1.2 in \cite{HP2}.
\begin{itemize}
\item[(1)] 
For $\lambda = 1$, we have $\ln_{\lambda} \Tr[ e_{\lambda}^{A+\ln_{\lambda}Y }] =  \Tr[X^{1-\lambda}A] -D_{\lambda} (X\vert Y) $ by easy calculations with the usual convention $X^0=I$. 
Thus we assume $\lambda \in (0,1)$.
Let us denote
$$F_{\lambda}(X) = \Tr[X^{1-\lambda} A] -D_{\lambda} (X\vert Y) $$
for a nonnegative matrix $X$ with $\Tr[X]=1$. If we take the Schatten decomposition $X = \sum_{j=1}^n r_j E_j $, where all
$E_j$, $(j=1,2,\cdots ,n)$ are projections of rank one with $\sum_{j=1}^n E_j =I$ and $r_j \geq 0$, $(j=1,2,\cdots ,n)$ with $\sum_{j=1}^n r_j =1$, then we rewrite
$$F_{\lambda}\left(\sum_{j=1}^n r_j E_j\right) = \sum_{j=1}^n \left\{ r_j^{1-\lambda}\Tr[E_j A] + \frac{1}{\lambda} r_j^{1-\lambda}  \Tr[E_jY^{\lambda}] - \frac{1}{\lambda} r_j \Tr[E_j]\right\}.$$
Since we have 
$$ \frac{\partial}{\partial r_j} F_{\lambda} \left( \sum_{j=1}^n r_j E_j\right) \vert _{r_j =0} = +\infty$$ 
%(which means $F_{\lambda}$ is concave function),  because $f$: concave iff $f(x) \leq f(c) +f'(c)(x-c),\,\, (a\leq x \leq b)$),
$F_{\lambda}(X)$ attains its maximum at a nonnegative matrix $X_0$ with $\Tr[X_0] = 1$. Then for any Hermitian matrix $S$ with $\Tr[S] =0$ (since $\Tr[X_0 + t S]$ must be $1$), we have
$$0 = \frac{d}{dt} F_{\lambda}(X_0+tS) \vert _{t=0} = (1-\lambda) \Tr[S (X_0^{-\lambda}A +\frac{1}{\lambda} X_0^{-\lambda}Y^{\lambda} )],$$
 so that $X_0^{-\lambda}A +\frac{1}{\lambda} X_0^{-\lambda}Y^{\lambda} = c I$ for $c \in \mathbf{R}$. Thus we have 
$$ X_0 = \frac{e_{\lambda}^{A+\ln_{\lambda}Y}}{\Tr[e_{\lambda}^{A+\ln_{\lambda}Y}]}$$  
by the normalization condition. By the formulae $\ln_{\lambda}\frac{y}{x} = \ln_{\lambda}y + y^{\lambda} \ln_{\lambda} \frac{1}{x}$ 
and $\ln_{\lambda} \frac{1}{x} = -x^{-\lambda} \ln_{\lambda}x$, we have
$$F_{\lambda}(X_0) = \ln_{\lambda}\Tr[e_{\lambda}^{A+\ln_{\lambda}Y}]. $$
\item[(2)] For $\lambda = 1$, we have $D_{\lambda}(X\vert e_{\lambda}^B)  =  \Tr[X^{1-\lambda} A] - \ln_{\lambda} \Tr[e_{\lambda}^{A+B}] $ 
 by easy calculations. Thus we assume $\lambda \in (0,1)$.
It follows from (1) that the functional $g(A) = \ln_{\lambda} \Tr[e_{\lambda}^{A+B}]$ 
defined on the set of all nonnegative matrices is convex, due to triangle inequality on $\max$. 
Now let $A_0 = \ln_{\lambda} X -B$, and denote
$$G_{\lambda}(A) = \Tr[X^{1-\lambda}A] - \ln_{\lambda}\Tr[e_{\lambda}^{A+B}], $$
which is concave on the set of all nonnegative matrices. Then for any nonnegative matrix $S$, there exists a nonnegative matrix $A_0$ such that
$$\frac{d}{dt}G_{\lambda}(A_0+tS) \vert _{t=0} = 0. $$
Therefore $G_{\lambda}(A)$ attaines the maximum $G_{\lambda}(A_0) = D_{\lambda}(X\vert e_{\lambda}^B)$.
\end{itemize}
\hfill \qed
\vspace*{2mm}

Taking the limit as $\lambda \to 0$, Theorem \ref{gen_rela_vari_exp} recovers the similar form of Lemma 1.2 in \cite{HP2} under the assumption of nonnegativitiy of $A$.
If $Y=I$ and $B=0$ in (1) and (2) of Theorem \ref{gen_rela_vari_exp}, respectively, then we obtain the following corollary.

\begin{Cor}\label{gen_vari_exp}
\begin{itemize}
\item[(1)] If $A$ is a nonnegative matrix, then
$$\ln_{\lambda}\Tr[e_{\lambda}^A] = \max \left\{ \Tr[X^{1-\lambda}A] + S_{\lambda}(X) : X \geq 0,\, \Tr[X]=1\right\}.$$
\item[(2)] For a density matrix $X$, we have
$$-S_{\lambda}(X) = \max \left\{ \Tr[X^{1-\lambda}A] -\ln_{\lambda}\Tr[e_{\lambda}^A] : A \geq 0 \right\}. $$
\end{itemize}
\end{Cor}
Taking the limit as $\lambda \to 0$, Corollary \ref{gen_vari_exp} recovers the similar form of Theorem 1 in \cite{BPL} under the assumption of nonnegativitiy of $A$.

%\begin{Rem}
%Form (1) of Corollary \ref{gen_vari_exp}, we may regard the functional $f_{\lambda} (X) \equiv \Tr[X^{1-\lambda}A] + S_{\lambda}(X)$
%depending on $\lambda \in (0,1]$ 
%as a generalized free energy functional of the quantum state $X$.
%\end{Rem}

\section{Generalized logarithmic trace inequalities}
In this section, we derive some trace inequalities in terms of the results obtained in the previous section.
From (1) of Corollary \ref{gen_vari_exp}, we have the generalized thermodynamic inequality:
\begin{equation}  \label{gen_Thermo_ineq}
\ln_{\lambda}\Tr[e_{\lambda}^H] \geq \Tr[D^{1-\lambda}H] + S_{\lambda}(D), 
\end{equation}
for a density matrix $D$ and a nonnegative matrix $H$.
Putting $D = \frac{A}{\Tr[A]}$ and $H = \ln_{\lambda}B$ in Eq.(\ref{gen_Thermo_ineq}) for $A \geq 0$ and $B \geq I$, 
we have the generalized Peierls-Bogoliubov inequality (cf.Theorem 3.3 of \cite{FYK}):
\begin{equation}  \label{gen_Pei_Bog}
(\Tr[A])^{1-\lambda} \left( \ln_{\lambda}\Tr[A]  - \ln_{\lambda} \Tr[B] \right) \leq  \Tr[A^{1-\lambda} (\ln_{\lambda}A - \ln_{\lambda} B) ],
\end{equation}
for nonnegative matrices $A$ and $B \geq I$.

\begin{Lem} \label{vari_exp_rela_lemma}
The following statements are equivalent.
\begin{itemize}
\item[(1)] $F_{\lambda}(A) = \ln_{\lambda} \Tr[e_{\lambda}^A]$ is convex in a Hermitian matrix $A$.
\item[(2)] $f_{\lambda}(t) = \ln_{\lambda} \Tr[e_{\lambda}^{A+tB}] $ is convex in $t\in \hbox{R}$.
\end{itemize}
\end{Lem}
{\it Proof}:
 Putting $A_1 = A+ x B$ and $A_2 = A + y B$ in
$$\ln_{\lambda} \Tr[e_{\lambda}^{\mu A_1+(1-\mu_2) A_2}] \leq \mu \ln_{\lambda} \Tr[e_{\lambda}^{A_1}] +(1-\mu) \ln_{\lambda}\Tr[e_{\lambda}^{A_2}], $$
we find the convexity of $f_{\lambda}(t)$. Thus (1) implies (2).

Conversely, putting $A = A_2$, $B= A_1 - A_2$ and $x=1,y=0$ in
$$\ln_{\lambda}\Tr[e_{\lambda}^{A + (\mu x +(1-\mu)y )B}] \leq \mu \ln_{\lambda}\Tr[e_{\lambda}^{A + x B}] + (1-\mu) \ln_{\lambda} \Tr[e_{\lambda}^{A + yB}], $$
we find the convexity of $F_{\lambda}(A)$. Thus (2) implies (1).

\hfill \qed
\vspace*{2mm}

\begin{Cor}  \label{corollary_trace_inequality}
For nonnegative matrices $A$ and $B$, we have
\begin{equation}  \label{cor_from_bogo_ineq}
\ln_{\lambda}\Tr[e_{\lambda}^{A+B}] - \ln_{\lambda}\Tr[e_{\lambda}^A] \geq \frac{\Tr[B (e_{\lambda}^A)^{1-\lambda}]}{(\Tr[e_{\lambda}^A])^{1-\lambda}}. 
\end{equation}
\end{Cor}
{\it Proof}:
From Theorem \ref{gen_rela_vari_exp}, $F_{\lambda}(A)=\ln_{\lambda}\Tr[e_{\lambda}^A]$ is convex. Due to Lemma \ref{vari_exp_rela_lemma},
$f_{\lambda}(t) = \ln_{\lambda} \Tr[e_{\lambda}^{A+tB}]$ is convex, which implies that $f_{\lambda}(1) \geq f_{\lambda}(0) +f'_{\lambda}(0)$. 
Thus we have the corollary by simple calculations.  
\hfill \qed
\vspace*{2mm}

%Note that we also obtain the inequality (\ref{cor_from_bogo_ineq}) by putting $A = e_{\lambda}^H$ and $B = e_{\lambda}^{H+K}$ in the inequality (\ref{gen_Pei_Bog}).
%In addition, putting $A=\ln_{\lambda}D$ and $B= H -\ln_{\lambda}D$ for a density matrix $D$ in the inequality (\ref{cor_from_bogo_ineq}),
% we obtain the inequality (\ref{gen_Thermo_ineq}).
%Thus we have the following proposition.

%\begin{Prop}  \label{prop_tra_ineq_gen}
%The following conditions are equivalent.
%\begin{itemize}
%\item[(1)] The generalized thermodynamic inequality (\ref{gen_Thermo_ineq}).
%\item[(2)] The generalized Peierls-Bogoliubov inequality (\ref{gen_Pei_Bog}).
%\item[(3)] The trace inequality (\ref{cor_from_bogo_ineq}) given in Corollary \ref{corollary_trace_inequality}.
%\end{itemize}
%\end{Prop}
%Taking the limit as $\lambda \to 0$, Proposition \ref{prop_tra_ineq_gen} recovers Theorem 2 in \cite{BPL}.

\section{Generalized exponential trace inequality}
For nonnegative real numbers $x,y$ and $0< \lambda \leq 1$, the relations $e_{\lambda}^{x+y} \leq e_{\lambda}^{x+y+\lambda xy} =e_{\lambda}^{x} e_{\lambda}^{y}$ hold.
These relations naturally motivate us to consider the following inequalities in the noncommutative case.

\begin{Prop} \label{GT_ineq_gen_before}
For nonnegative matrices $X$ and $Y$, and $0 < \lambda \leq 1$, we have
$$
\Tr[  e_{\lambda}^{X+Y}  ] \leq \Tr[  e_{\lambda}^{X+Y+\lambda Y^{1/2}XY^{1/2}}  ].
$$
\end{Prop}
{\bf (Proof)}
 Since $Y^{1/2}XY^{1/2} \geq 0$ for nonnegative matrices $X$ and $Y$, we have
$$ 
I+\lambda (X+Y) \leq I +\lambda \left\{X+Y + \lambda (Y^{1/2}XY^{1/2})\right\}.
$$
It is known \cite{Petz1} that $A \leq B$ implies $\Tr[f(A)] \leq \Tr[f(B)]$, if $A$ and $B$ are Hermitian matrices and $f:\hbox{R} \to \hbox{R}$ is an increasing function.
Thus we have 
$$
\Tr[\left(I+\lambda (X+Y)\right)^{1/\lambda}] \leq \Tr[\left(I +\lambda \left\{X+Y + \lambda (Y^{1/2}XY^{1/2})\right\}\right)^{1/\lambda}].
$$
for $0 < \lambda \leq 1$.
\hfill \qed 
\vspace*{2mm}

Note that we have the matrix inequality :
$$
e_{\lambda}^{X+Y}  \leq e_{\lambda}^{X+Y+\lambda Y^{1/2}XY^{1/2}}  
$$
for $\lambda \geq 1$ by the application of the L\"owner-Heinz inequality \cite{Low,Hei,Ped}.

\begin{Prop}  \label{GT_ineq_gen}
For nonnegative matrices $X,Y$, and $\lambda \in (0,1]$, we have
\begin{equation}   \label{GT_ineq_gen_EQ}
\Tr[e_{\lambda}^{X+Y+\lambda XY}] \leq \Tr[e_{\lambda}^X e_{\lambda}^Y].
\end{equation}
\end{Prop}
{\it Proof}:
For $\lambda \in (0,1] $, we have 
\begin{equation}  \label{mod_lieb_thirring}
\Tr[(AB)^{1/\lambda}] \leq \Tr[A^{1/\lambda} B^{1/\lambda}]
\end{equation}
from the special case of the Lieb-Thirring inequality \cite{LT} (see also \cite{HH,WZ}), 
$\Tr[(AB)^{\alpha}] \leq \Tr[A^{\alpha}B^{\alpha}]$ for nonnegative matrices $A$, $B$ and any real number $\alpha \geq 1$.
Putting $A= I +\lambda X$ and $B=I +\lambda Y$ in Eq.(\ref{mod_lieb_thirring}), 
we have 
$$
\Tr[\left\{ (I +\lambda X)(I +\lambda Y)\right\}^{1/\lambda}]  \leq \Tr[  (I +\lambda X)^{1/\lambda} (I +\lambda Y)^{1/\lambda}],
$$
which is the desired one.
\hfill \qed 
\vspace*{2mm}

Notice that Golden-Thompson inequality \cite{Gol,Tho}, 
$$\Tr[e^{X+Y}] \leq \Tr[e^X e^Y]$$
which holds for Hermitian matrices $X$ and $Y$, is recovered by taking the limit as $\lambda \to 0$ in Proposition \ref{GT_ineq_gen},
in particular case of nonnegative matrices $X$ and $Y$.

\section{Concluding remarks}

Since $\Tr[HZHZ] \leq \Tr[H^2Z^2]$ for Hermitian matrices $H$ and $Z$ \cite{Petz2,Fujii}, we have for nonnegative matrices $X$ and $Y$,
$$\Tr[(I+X+Y+Y^{1/2}XY^{1/2})^2] \leq \Tr[(I+X+Y+XY)^2]$$
 by easy calculations. This implies the inequality
$$ \Tr[e_{1/2}^{X+Y+1/2Y^{1/2}XY^{1/2}}] \leq \Tr[e_{1/2}^{X+Y+1/2XY}].$$
 Thus we have 
\begin{equation}   \label{eq_1_2_GT_gen}
\Tr[e_{1/2}^{X+Y}] \leq \Tr[e_{1/2}^Xe_{1/2}^Y]
\end{equation}
from Proposition \ref{GT_ineq_gen_before} and Proposition \ref{GT_ineq_gen}. 
Putting $B=\ln_{1/2}Y$ and $A=\ln_{1/2}Y^{-1/2}XY^{-1/2}$ in (2) of Theorem \ref{gen_rela_vari_exp} 
under the assumption of $I\leq Y\leq X$ and using Eq.(\ref{eq_1_2_GT_gen}), 
we have
\begin{eqnarray}
D_{1/2}(X\vert Y)  &=& D_{1/2}(X\vert e_{1/2}^{\ln_{1/2}Y}) \nonumber \\
&\geq& \Tr[X^{1/2}A] -\ln_{1/2} \Tr[e_{1/2}^{A+B}] \nonumber \\
&\geq&  \Tr[X^{1/2}A] -\ln_{1/2} \Tr[  e_{1/2}^{A} e_{1/2}^{B}    ] \nonumber \\
&=& \Tr[X^{1/2}\ln_{1/2} Y^{-1/2}XY^{-1/2} ]  -\ln_{1/2} \Tr[Y^{-1/2}XY^{-1/2} Y] \nonumber \\
&=& \Tr[X^{1/2}\ln_{1/2} Y^{-1/2}XY^{-1/2} ],  \label{lower_bound_rel_ent_1_2}
\end{eqnarray}
which gives a lower bound of the Tsallis relative entropy in the case of $\lambda = 1/2$ and $I\leq Y\leq X$.

\section*{Acknowledgement}
We would like to thank the reviewers for providing valuable comments to improve the manuscript.
This work was supported by the Japanese Ministry of Education, Science, Sports and Culture, Grant-in-Aid for 
Encouragement of Young scientists (B), 17740068.

\end{document}